\documentclass[12pt]{article}
\usepackage{graphicx}
\usepackage{amssymb}
\usepackage{amscd}
\usepackage{amsmath}
\usepackage{cite}
\usepackage[british]{babel}
\usepackage{microtype}

\begin{document}

{\hbox to\hsize{\hfill January 2019 }}

\bigskip \vspace{3\baselineskip}

\begin{center}
{\bf \Large 
A brief remark on convexity of effective potentials and de Sitter Swampland conjectures }

\bigskip

\bigskip

{\bf Archil Kobakhidze \\ }

\smallskip

{ \small \it
School of Physics, The University of Sydney, NSW 2006, Australia, \\
E-mails: archil.kobakhidze@sydney.edu.au 
\\}

\bigskip
 
\bigskip

\bigskip

{\large \bf Abstract}

\end{center}
\noindent 
Recently proposed de Sitter Swampland conjectures imply non-trivial constraints  on  a scalar field potential in any effective field theory that admits a quantum gravity completion. The original conjecture apparently excludes many phenomenologically motivated scalar potentials with de Sitter extrema, such as the perturbative Standard Model Higgs potential and the QCD axion potential, as the viable low-energy theories. Subsequently, the refined, weaker conjecture was proposed.  However,  
the full effective potential, having been defined as a Legendre transform,  is necessarily convex. This ensures that the original Swampland conjecture can actually be satisfied in phenomenologically relevant models, providing no de Sitter vacua exist.      

\bigskip
 
\bigskip

\bigskip

Motivated by string theory considerations, it has been conjectured in  Ref. \cite{Obied:2018sgi} that that the potential $V(\phi_i)$ for scalar fields, $\phi_i$, in any low-energy effective theory that admits embedding into a fundamental theory with quantum gravity included, must satisfy the following universal bound:
\begin{equation}
\vert \partial_{\phi_i}V\vert \geq \frac{c}{M_p}\cdot V~, 
\label{1}
\end{equation}
for some positive constant $c>0$ of order 1. Here, $\vert \partial_{\phi_i}V\vert=\sqrt{\sum_i \left (\partial_{\phi_i}V\right)^2}$ computed in the basis of scalar fields with canonical kinetic terms and minimal coupling to gravity, and $M_p\approx 2.4\cdot 10^{18}$ GeV is the reduced Planck mass. Notably, the condition (\ref{1}) excludes de Sitter extrema for which $\partial_{\phi_i}V=0,~V>0$, and in particular de Sitter ground state as the origin of the observed accelerated expansion of the universe at present epoch. Hence the name, the de Sitter Swampland conjecture. Note, that the bound (\ref{1}) with $ c =\sqrt{54/13}$  has been obtained in \cite{Hertzberg:2007wc} in the context of type IIA string theory. Also, incompatibility of de Sitter vacua with quantum corpuscular description has been argued previously in \cite{Dvali:2013eja}. The relation between the corpuscular description and the conjecture (\ref{1}) has been illuminated in \cite{Dvali:2018jhn}. 

The original de Sitter Swampland conjecture of Eq. (\ref{1}) appeared to be rather restrictive for cosmological \cite{Agrawal:2018own} and particle physics \cite{Denef:2018etk, Murayama:2018lie, Choi:2018rze, Hamaguchi:2018vtv} applications. Subsequently, the weaker conjecture was proposed in \cite{Ooguri:2018wrx} (see also \cite{Garg:2018reu}), 
\begin{equation}
\vert \partial_{\phi_i}V\vert \geq \frac{c}{M_p}\cdot V~~\rm{or}~~\rm{min}\left( \partial_{\phi_i}\partial_{\phi_j}V\right) \leq -\frac{c'}{M^2_p}\cdot V~, 
\label{2}
\end{equation}  
where $c'$ is another positive constant of order 1 and  $\rm{min}\left( \partial_{\phi_i}\partial_{\phi_j}V\right)$ is a minimum eigenvalue of the Hessian matrix $\partial_{\phi_i}\partial_{\phi_j}V$. While the de Sitter minima are still prohibited by the refined conjecture, de Sitter maxima are now allowed, providing $\rm{min}\left( \partial_{\phi_i}\partial_{\phi_j}V\right)$ is sufficiently negative. 

However, if the potential $V$ is understood as the full effective potential, which is defined via a Legendre transform, it must be everywhere convex \cite{Symanzik:1969ek, Iliopoulos:1974ur}. The Hessian $\partial_{\phi_i}\partial_{\phi_j}V$ then is necessarily positive semi-definite  and hence the second inequality in Eq. (\ref{2}) becomes redundant. In other words, the strong de Sitter conjecture (\ref{1}) is automatically satisfied, providing the scalar potential admits no de Sitter minima. 

The standard perturbative 1PI effective potential fails to reflect the convexity property in a region between minima, because the perturbative calculations are based on the expansion around one definite vacuum configuration. Such constant field configuration, however, corresponds  to a unstable saddle point rather than a minimum in the spinodal region of the potential. Instead, one must take into account large-wavelength spinodal modes, an inhomogeneous mixture of minima. As a result, the effective potential $V_k$ computed at energy $k$ gets flattened in the regions between the minima as $k\to 0$ \cite{Weinberg:1987vp, Ringwald:1989dz}.  Hence,  large-scale nonperturbative fluctuations dominate near the region of the effective potential between the minima and are essential part of the consistent description of the low-energy effective theory. It is an interesting question whether these fluctuations are related to the tower of large-wavelength  states, which are discarded in \cite{Ooguri:2006in, Ooguri:2018wrx} from the spectrum of the effective theory. This may also have its bearing to the so-called field distance conjecture \cite{Ooguri:2006in}. 

To conclude, the often ignored large scale nonperturbative fluctuations of scalar fields flatten regions between minima of the effective potential and restore it convexity. Thus, these fluctuations are the essential part of the consistent low-energy description of the theory.   Therefore, the strong version of the de Sitter Swampland conjecture is automatically satisfied in phenomenologically relevant models, providing there are no de Sitter minima.

\paragraph{Acknowledgement} I acknowledge useful discussions on the Swampland conjectures with Mark Hertzberg and Tsutomu Yanagida. The work was partially supported by the Australian Research Council and the Rustaveli National Science Foundation.

\end{document}